\documentclass[10pt,prl,twocolumn,preprint,nofootinbib]{revtex4}

\usepackage{epsfig}
\usepackage{verbatim}
\usepackage{amsmath,amsfonts,amssymb}
\usepackage{t1enc}
\usepackage{color}

\providecommand{\openone}{\leavevmode\hbox{\small1\kern-3.8pt\normalsize1}}

\newcommand{\Vl}{V_L}
\newcommand{\Vr}{V_R}
\newcommand{\gl}{g_L}
\newcommand{\gr}{g_R}

\newcommand{\fz}{F_0}
\newcommand{\fl}{F_L}
\newcommand{\fr}{F_R}

\begin{document}

\title{New limits on anomalous contributions to the $Wtb$ vertex}

\author{
J.~L.~Birman$^{1}$,
F.~D\'eliot$^{2}$,
M.~C.~N.~Fiolhais$^{1,3,4}$,
A.~Onofre$^{5}$,
C.~M.~Pease$^{1}$
\\[3mm]
{\footnotesize {\it 
$^1$ Department of Physics, City College of the City University of New York, \\ 
     160 Convent Avenue, New York 10031, NY, USA \\
$^2$ Institute of Research into the Fundamental Laws of the Universe, CEA, Saclay, France \\
$^3$ Department of Physics, New York City College of Technology,\\ 300 Jay Street, Brooklyn, NY 11201, USA \\
$^4$ LIP, Departamento de F\'{\i}sica, Universidade de Coimbra, 3004-516 Coimbra, Portugal \\
$^5$ LIP, Departamento de F\'{\i}sica, Universidade do Minho, Campus de Gualtar, 4710-057 Braga, Portugal
}}
}

\begin{abstract}
The latest and most precise top quark measurements at the LHC and Tevatron are used to establish new limits on the $Wtb$ vertex. Recent results on the measurements of the $W$-boson helicity fractions and single top quark production cross section are combined in order to establish new limits at 95\%~CL (confidence level). The allowed regions for these limits are presented, for the first time, in three-dimensional graphics, for both real and imaginary components of the different anomalous couplings, providing a new perspective on the impact of the combination of different physics observables. These results are also combined with the prospected future measurement of the single top quark production cross section and $W$-boson helicity fractions at the LHC.

\end{abstract}


\maketitle

\section{Introduction}

The recent discovery of a new scalar particle, compatible with the Standard Model (SM) Higgs boson, by the ATLAS and CMS collaborations at the LHC~\cite{atlash,cmsh}, brought us perhaps the most important missing piece of the SM, ever since the discovery of the top quark at Fermilab, in 1995~\cite{top-CDF,top-D0}. With the LHC  operating at a center-of-mass energy of 13 TeV, this is an exciting epoch for particle physics. The next decades will feature an unprecedented quest for new physics beyond the SM, that began at the LHC, and will hopefully continue with a future generation of colliders~\cite{AguilarSaavedra:2012vh}. 

The electroweak precision measurements performed at LEP and Tevatron, and, more recently, the observation of the Higgs boson with properties compatible with the one predicted by the SM, constitute strong experimental support for the theory. This implies that any possible new interaction coupling to the SM has to exist above the electroweak symmetry breaking scale, and perhaps beyond the TeV scale. For these reasons, precision measurements of top quark and Higgs boson properties, in particular of their couplings, constitute a fundamental component of direct searches for new physics beyond the SM, as they are expected to be sensitive to new physics at higher scales.

Top quarks decay through the weak interaction, mostly to a W boson and bottom quark with a mean lifetime estimated to be $10^{-25}$s. Due to its brief lifetime, shorter than the hadronization time scale ($10^{-24}$s), the top quark decays before hadronization can take place. The top quark spin information is, therefore, preserved by the decay products. By measuring the angular distributions of these decay products, it is possible to access the spin information and hence to probe the nature of the $Wtb$ vertex. In the SM, the $Wtb$ vertex has a (V-A) structure, which can be tested by measuring helicity fractions of the $W$-bosons produced in top quark decays, and/or the observables which depend on these helicity fractions~\cite{Kane:1991bg,delAguila:2002nf, AguilarSaavedra:2006fy}. Above the electroweak symmetry breaking scale, new physics effects can be parameterized in terms of an effective field theory approach~\cite{Buchmuller:1985jz}, where the most general $Wtb$ vertex can be expressed as~\cite{AguilarSaavedra:2008zc}:
\begin{eqnarray}
\mathcal{L}_{Wtb} & = & - \frac{g}{\sqrt 2} \bar b \, \gamma^{\mu} \left( \Vl
P_L + \Vr P_R
\right) t\; W_\mu^- \nonumber \\
& - &  \frac{g}{\sqrt 2} \bar b \, \frac{i \sigma^{\mu \nu} q_\nu}{M_W}
\left( \gl P_L + \gr P_R \right) t\; W_\mu^- + \mathrm{h.c.}  
\label{ec:lagr}
\end{eqnarray}
The coupling $\Vl = V_{tb} \simeq 1$, at tree level in the SM, while all the other couplings, $\Vr$, $\gl$, $\gr$ are dimensionless and complex, and equal to zero at tree level in the SM. Even though the anomalous couplings are absent in the SM at tree level, they may receive significant contributions from new physics effects beyond the SM (BSM). These contributions can be tested in top quark decays by measuring the $W$-boson helicity fractions, together with single top quark production measurements, and used to set limits on the anomalous couplings~\cite{Boos:1999dd, Najafabadi:2008pb, AguilarSaavedra:2008gt,Chen:2005vr, AguilarSaavedra:2010nx, Zhang:2010dr}.

In this paper, the most precise measurements of the single top quark production cross section at different centre-of-mass energies at the LHC and Tevatron~\cite{CMS:2016ayb,st-CMS,Chatrchyan:2012ep,Aaltonen:2015cra,Aad:2015eto,Chatrchyan:2012zca,Aad:2015upn,CDF:2014uma} are used to set limits on the real and imaginary components of the anomalous contributions to the $Wtb$ vertex. These results are also combined with recent LHC results obtained by CMS~\cite{Wh-CMS} for the $W$-boson helicity fractions, which together with the ATLAS~\cite{Wh-ATLAS} and Tevatron results~\cite{whel-CDF_D0}, are in good agreement with next-to-next-to-leading-order (NNLO) SM predictions~\cite{Czarnecki:2010gb}. The combination of all the measurements described above, and the estimation of limits, is performed with the {\sc TopFit}~\cite{topfit} program by taking the global uncertainty of each measurement and their correlations, whenever known. {It must be stressed that this combination of measurements is a phenomenological exercise, and not an official combination from the experiments.} The allowed regions are presented for both real and imaginary components of the anomalous couplings, at 95\% confidence level (CL), in two-dimensional and three dimensional representations, allowing the visualization of the dependences and correlations between three different anomalous couplings at a time. These results present a different approach which signifcantly differs from the one commonly used in the literature, \emph{i.e.} all real and imaginary parts of the new couplings are kept as unconstrained free parameters of the global fit. Moreover, these results also present a significant improvement when combined with the prospected results for the $t$-channel single top cross section and $W$-boson helicity fractions measurements in future LHC runs at 14 TeV with a luminosity of 3000~fb$^{-1}$~\cite{Schoenrock:2013jka}.

\section{Experimental Results}

The $t$-channel single top quark production cross section has been measured with great precision at different center-of-mass energies, both at the LHC and Tevatron.  
CMS recently measured the t-channel single top-quark inclusive cross section using proton-proton collisions at 13 TeV~\cite{CMS:2016ayb}, corresponding to an integrated luminosity of 2.3~fb$^{-1}$, with one muon in the final state: 
\begin{align}
& \sigma_{t-ch}^{(13~\textrm{TeV})} = 227.8^{+33.7}_{-33.0}~{\rm pb}\, .
\end{align}
The same measurement had already been performed by CMS at center-of-mass energy of 8 TeV, using a total integrated luminosity of 19.7~fb$^{-1}$~\cite{st-CMS},
\begin{align}
& \sigma_{t-chan}^{(8~\textrm{TeV})} = 83.6 \pm 2.3\;\text{(stat)} \pm 7.4\;\text{(syst)} ~{\rm pb}\,.
\end{align}
The cross section was measured inclusively, in final states with a muon or an electron. Previous results from CMS at 7~TeV were also taken into account, with the single top quark production cross section in the $t$-channel measured to be~\cite{Chatrchyan:2012ep},
\begin{align}
& \sigma_{t-chan}^{(7~\textrm{TeV})} = 67.2 \pm 6.1~{\rm pb}\,.
\end{align}

Results from Tevatron were also used in this paper, in particular, the final combination of CDF and D0 measurements of cross sections for single-top-quark production in proton-antiproton collisions at a center-of-mass energy of 1.96 TeV. With a total integrated luminosity of up to 9.7 fb$^{-1}$ per experiment, the measured $t$-channel cross section is~\cite{Aaltonen:2015cra},
\begin{align}
& \sigma_{t-chan}^{(1.96~\textrm{TeV})} = 2.25^{+0.29}_{-0.31}~{\rm pb} \, .
\end{align}

The impact of the $Wt$ associated production was also considered in this paper. In particular, the measurement of the cross-section, with integrated luminosity of 20.3~fb$^{-1}$ of proton-proton collisions at 8 TeV, recorded by the ATLAS experiment~\cite{Aad:2015eto}, 
\begin{equation}
\sigma_{Wt} = 23.0 \pm 1.3\;\text{(stat)}^{+3.2}_{-3.5}\;\text{(syst)} \pm 1.1 \;\text{(lumi)} ~{\rm pb} \, ,
\end{equation}
was used. Results for the associated production of a single top quark and W boson in pp collisions at 7~TeV with the CMS experiment~\cite{Chatrchyan:2012zca},
\begin{equation}
\sigma_{Wt} = 16^{+5}_{-4}~{\rm pb} \, ,
\end{equation}
were also used. Moreover, the recent measurement of the $s$-channel single top-quark production cross-section, performed by ATLAS, was also taken into account. This measurement was achieved using proton-proton collisions at a centre-of-mass energy of 8 TeV, leading to an observed signal significance of 3.2 standard deviations~\cite{Aad:2015upn}, compatible with the SM, with a cross-section of 
\begin{equation}
\sigma_{s-chan} = 4.8 \pm 0.8 \;\text{(stat)} ^{+1.6}_{-1.3}\;\text{(sys)}~{\rm pb} \, . 
\end{equation}
The results for the single-top-quark production in the $s$-channel through the combination of the CDF and D0 measurements of the cross section in proton-antiproton collisions at a center-of-mass energy of 1.96 TeV~\cite{CDF:2014uma},
\begin{equation}
\sigma_{s-chan} = 1.29^{+0.26}_{-0.24}~{\rm pb} \, ,
\end{equation}
were also used. All measurements of single top quark production cross sections were assumed to be uncorrelated.

The currently most precise $W$-boson helicity fractions were measured in $t\bar t$ decays into the muon+jets channel, using a sample of events from proton-proton collisions at a centre-of-mass energy of 8 TeV, collected in 2012 with the CMS detector at the LHC. The measured $W$-boson helicity fractions are, 
respectively~\cite{Wh-CMS},
\begin{align}
& \fz = \phantom{-}0.659 \pm 0.015\;\text{(stat)} \pm 0.023\;\text{(syst)} \,, \notag \\
& \fl = \phantom{-}0.350 \pm 0.010\;\text{(stat)} \pm 0.024\;\text{(syst)} \,, \notag \\
& F_R = -0.009 \pm 0.006\;\text{(stat)} \pm 0.020\;\text{(syst)} \,.
\end{align}
Only $\fz$ and $\fl$ were used in the global fit with a correlation coefficient of \mbox{$\rho = -0.95$} for the LHC measurements~\cite{Bernardo:2014vha}. Tevatron results $\fz$ and $\fr$ were also taken into account, with a correlation coefficient of \mbox{$\rho = -0.86$}~\cite{Aaltonen:2012rz}.

Finally, the expected results for the $t$-channel single top quark production measurement in future LHC runs at 14~TeV and 33~TeV are also taken into account. In particular, in the single top $t$-channel final state with one lepton and a neutrino coming from the top quark decay, plus two jets (with one of them being tagged as a $b$-jet), the estimated cross-section precision is expected to reach 3.8\% for 3000~fb$^{-1}$ of 14 TeV $pp$ collision data~\cite{Schoenrock:2013jka}. In this scenario, the uncertainty on the $W$-helicity fractions was estimated to be half the current most precise measurement.

\section{Limits on anomalous couplings}

The dependences of the $Wtb$ helicity fractions and single top quark production cross sections on the anomalous couplings were derived in~\cite{AguilarSaavedra:2006fy} and~\cite{AguilarSaavedra:2008gt}, respectively. In this paper, these expressions were used to set limits on the anomalous contributions with the {\sc TopFit} program, which provides a proper treatment in establishing the allowed regions for these couplings, by taking into account different correlations between observables and global uncertainties.
No consideration was given to four-fermion contributions to the $t$-channel single top quark production cross section ~\cite{Cao:2007ea, AguilarSaavedra:2010zi}, and also no correlation was assumed between the helicities and the cross section measurements. The total uncertainty of each measurement is defined by adding in quadrature the corresponding statistical and systematic uncertainties.

\begin{figure*}[h]
\begin{center}
\includegraphics[height=8.5cm]{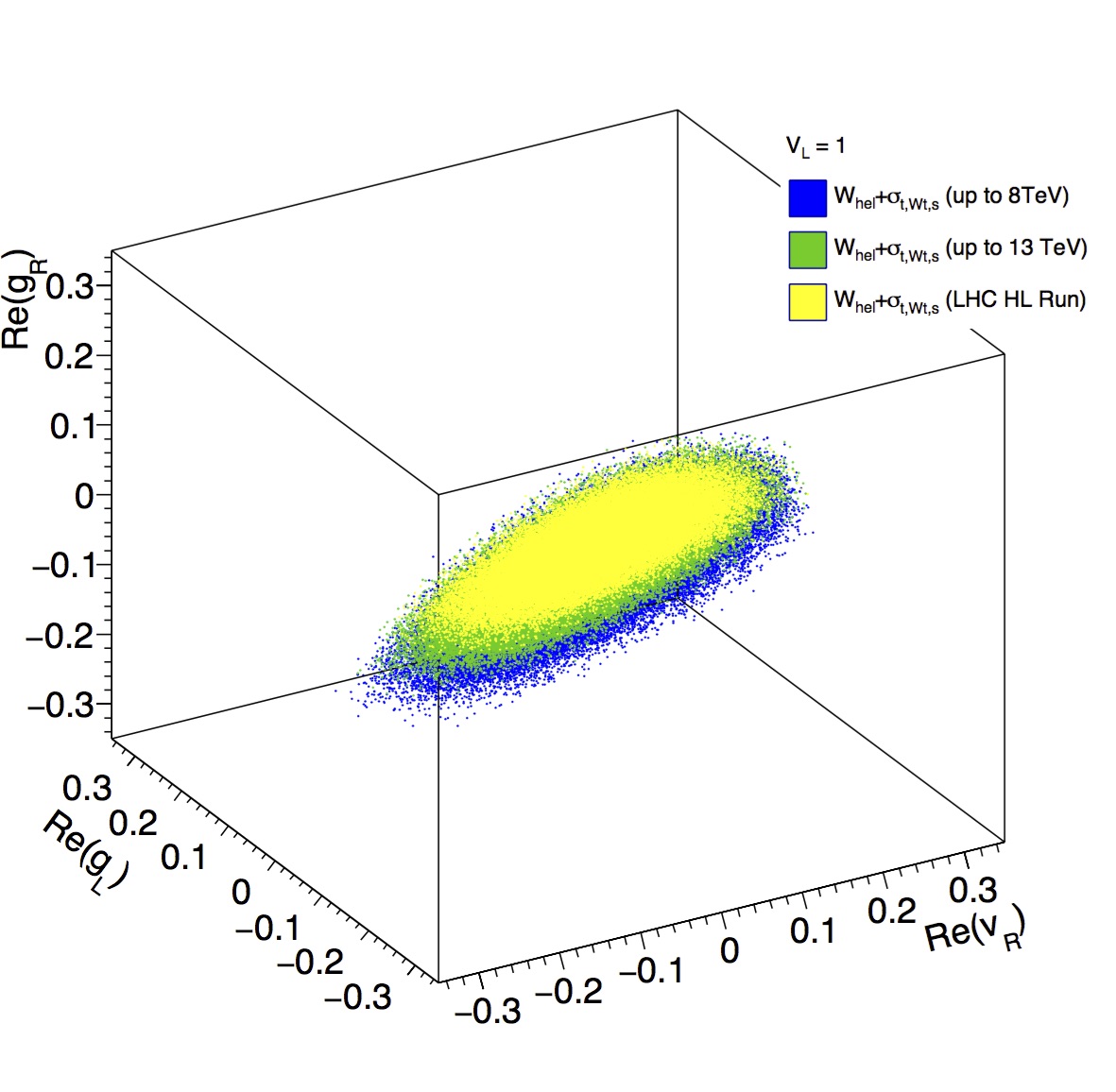}
\includegraphics[height=8.5cm]{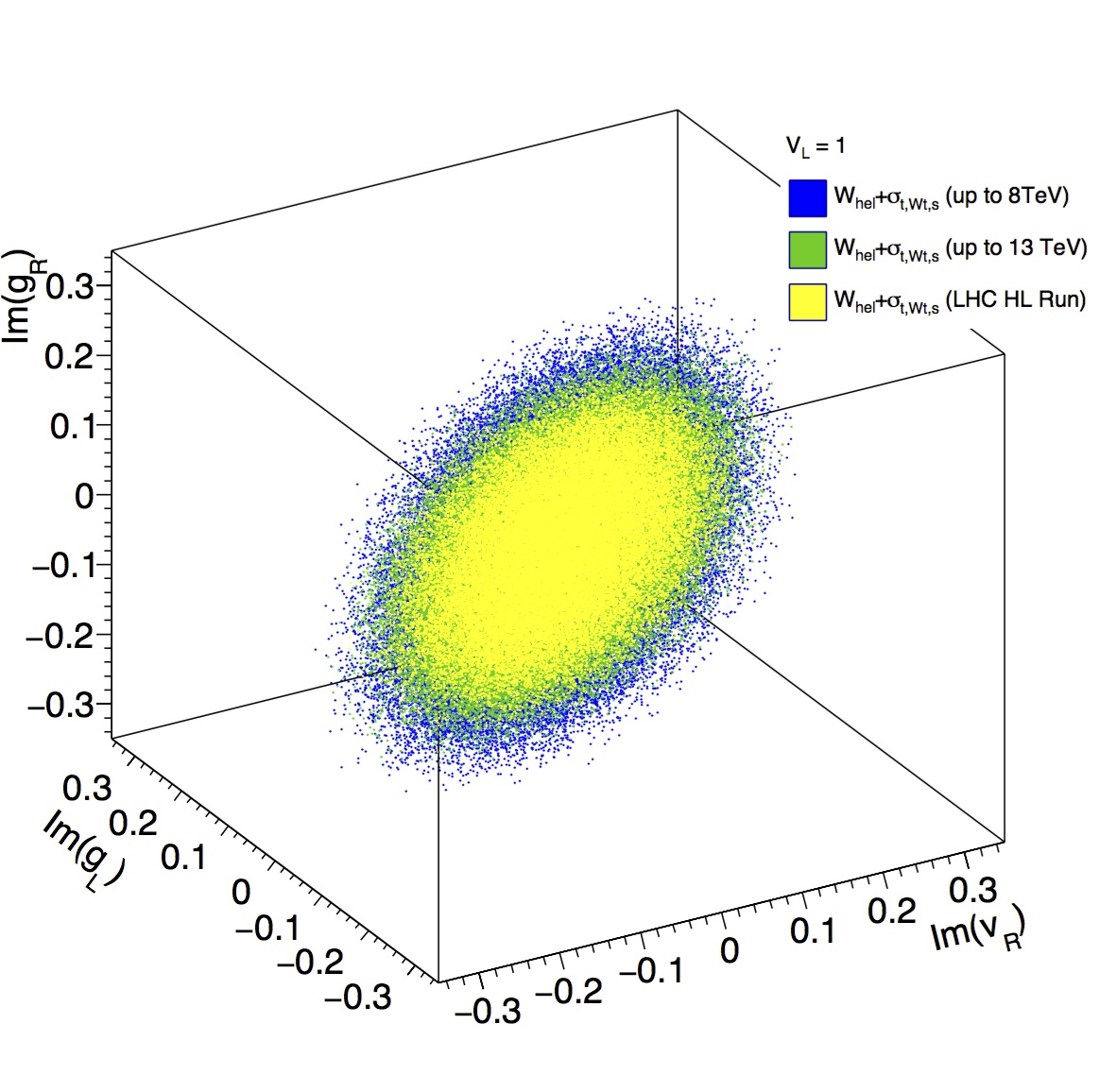}
\caption{Allowed region in $Re(\gr), Re(\gl)$ and $Re(\Vr)$ at 95\%~CL (left). Allowed region in $Im(\gr), Im(\gl)$ and $Im(\Vr)$ at 95\%~CL (right).}
\label{fig:lim1}
\end{center}
\end{figure*}

\begin{figure*}[h]
\begin{center}
\includegraphics[height=8.5cm]{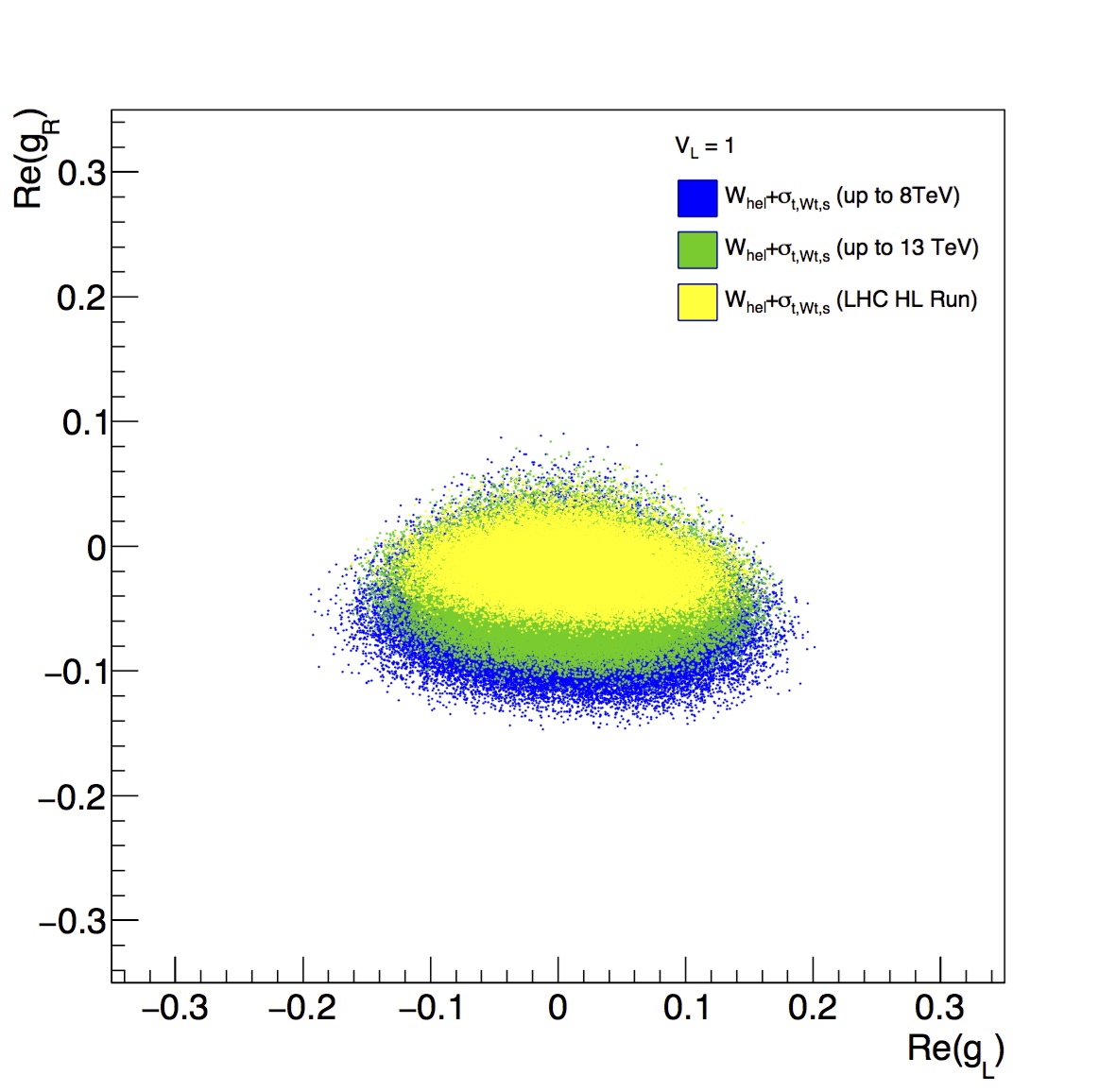} 
\includegraphics[height=8.5cm]{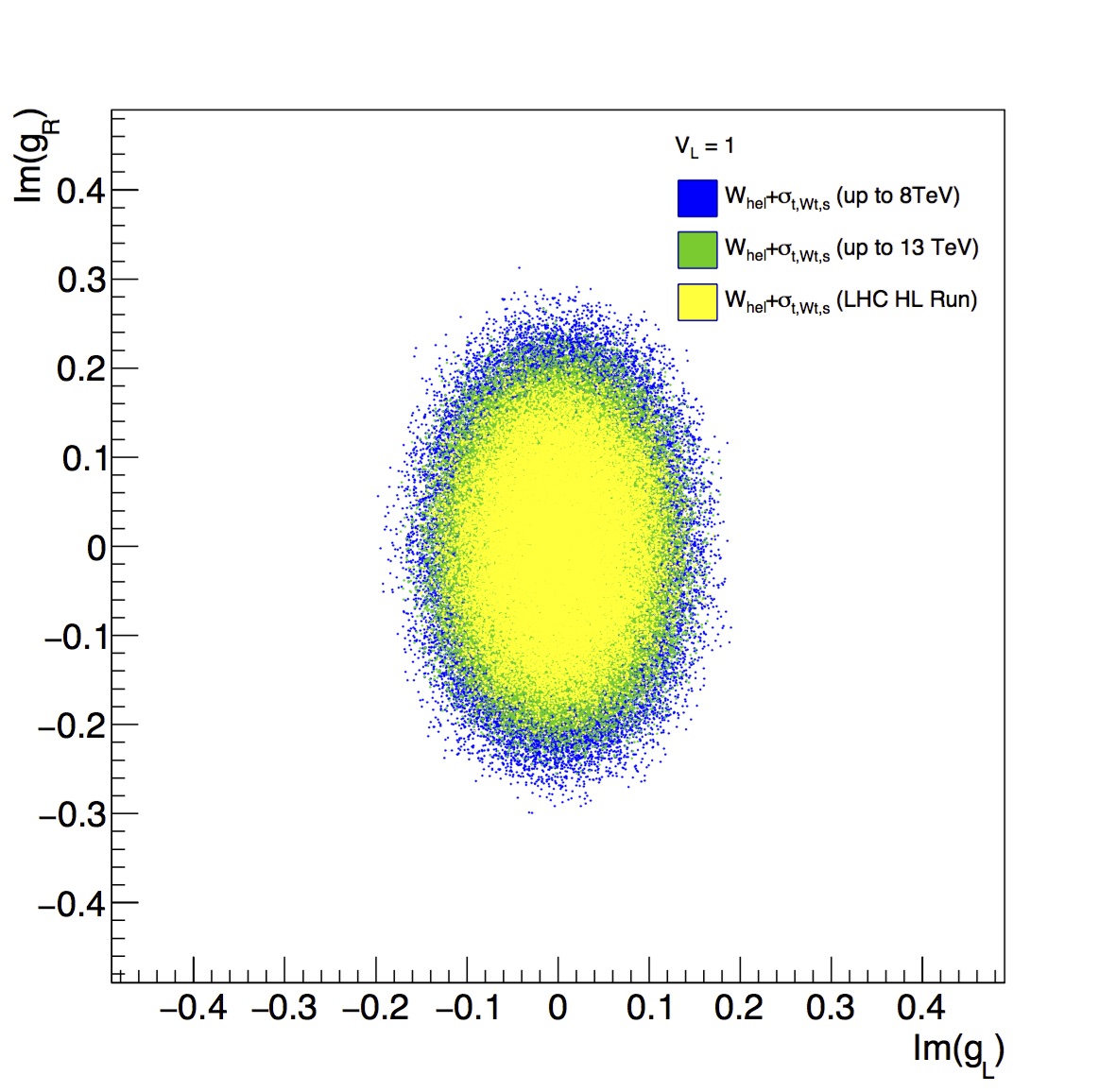}
\caption{95\%~CL allowed regions for the $\gl$ and $\gr$ complex couplings. The left plot corresponds to the real components, and the right plot corresponds to the imaginary ones.} 
\label{fig:lim3}
\end{center}
\end{figure*}

\begin{figure*}[h]
\begin{center}
\includegraphics[height=8.5cm]{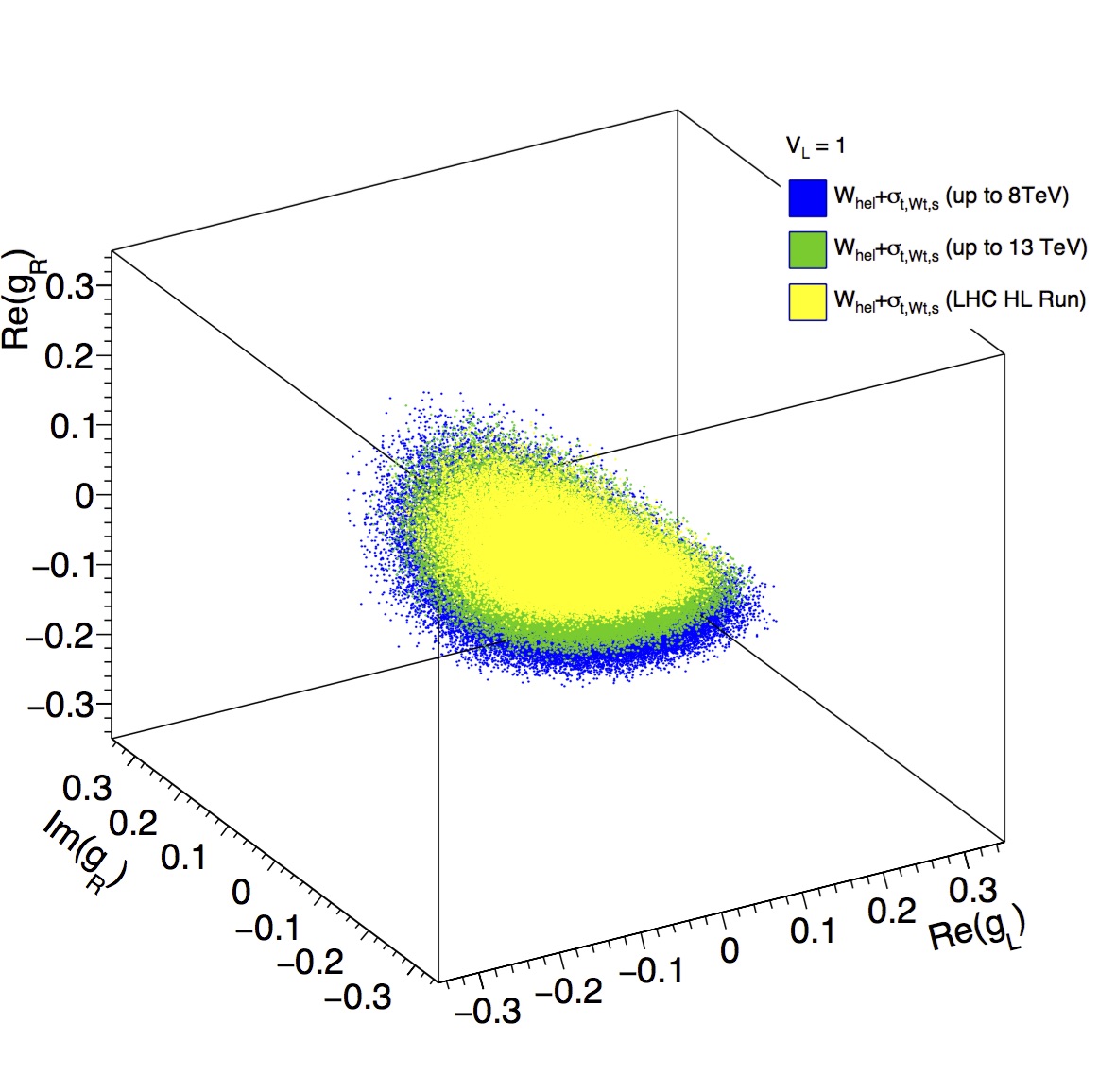}
\includegraphics[height=8.5cm]{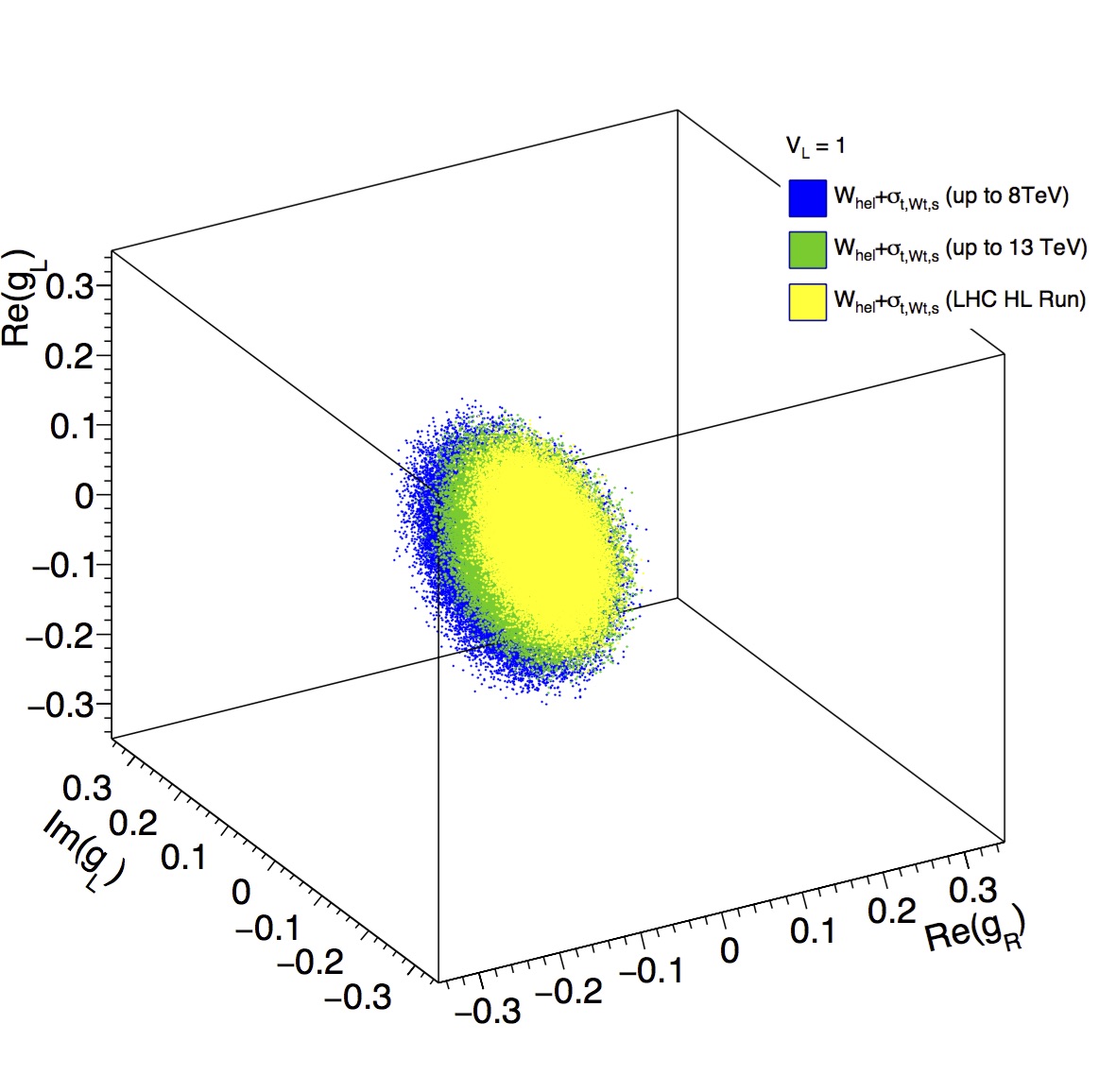}
\caption{Allowed region in $Re(\gr), Im(\gr)$ and $Re(\gl)$ at 95\%~CL (left). Allowed region in $Re(\gl), Im(\gl)$ and $Re(\gr)$ at 95\%~CL (right).}
\label{fig:lim4}
\end{center}
\end{figure*}

\begin{figure*}[h]
\begin{center}
\includegraphics[height=8.5cm]{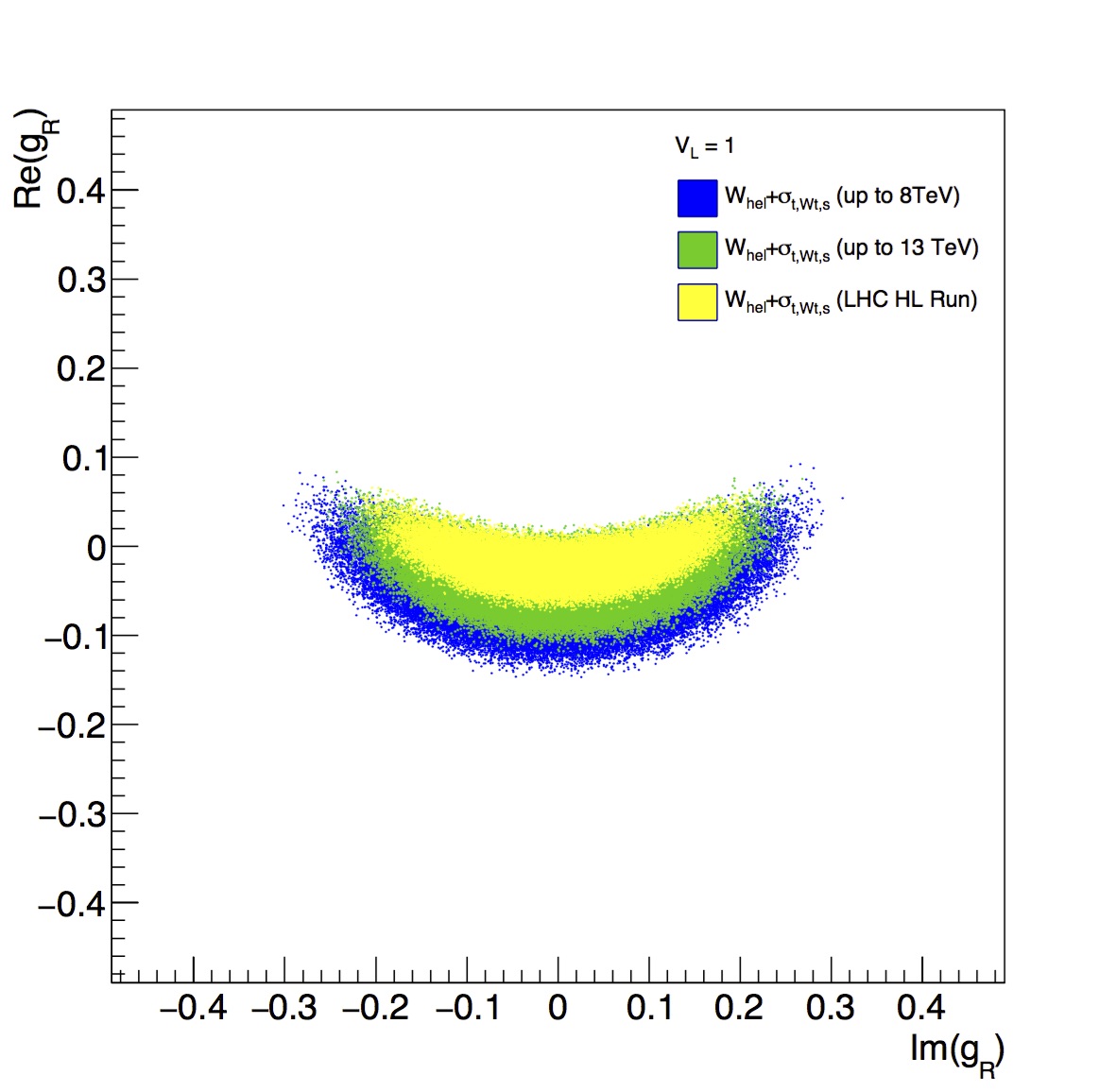} 
\includegraphics[height=8.5cm]{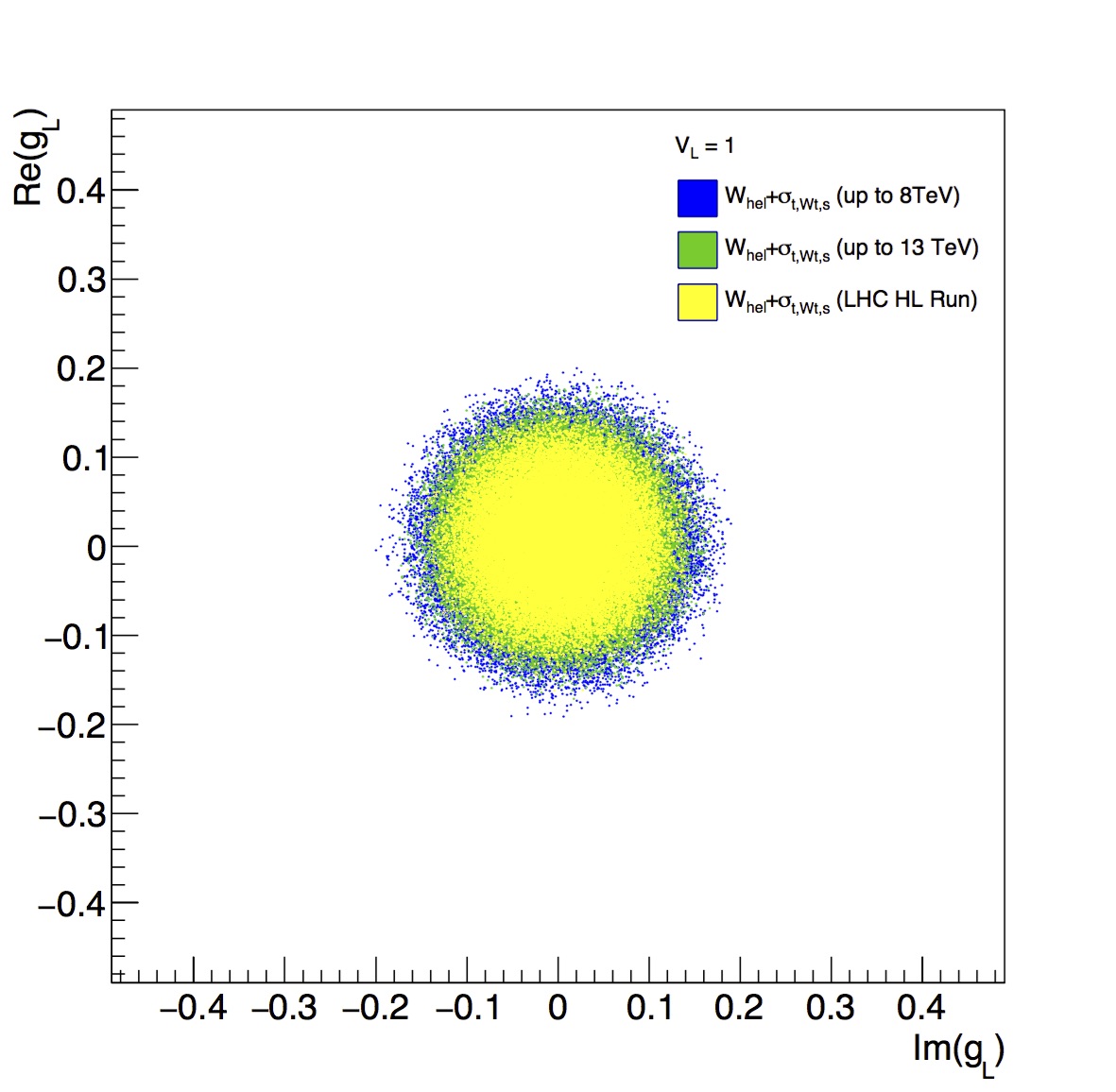}
\caption{95\%~CL allowed regions for the $\gl$ and $\gr$ complex couplings. The left plot corresponds to the real components, and the right plot corresponds to the imaginary ones.} 
\label{fig:lim5}
\end{center}
\end{figure*}

The allowed regions of phase-space are presented in three-dimensional plots of real and imaginary components of anomalous couplings $\Vr$, $\gl$, and $\gr$. In each case, the limits are set by {\sc TopFit} using top quark, $W$-boson, and bottom quark masses of $m_t = 175$ GeV, $M_W = 80.4$ GeV, and $m_b = 4.8$ GeV. Both real and imaginary components of the complex couplings are assumed to be non-vanishing, and limits are displayed for 95\%~CL. In each plot, from Figure~\ref{fig:lim1} to~\ref{fig:lim5}, use of different colors denotes the inclusion of additional measurements. A clear constriction of the volume occurs with each new consideration. The blue regions correspond to the 95\%~CL limits using the previously most precise $W$-boson helicity fractions and single top quark production measurements from LHC (up to 8~TeV), together with all Tevatron results ($W$-boson helicity fractions and all single top quark production cross-sections), as discussed in~\cite{Bernardo:2014vha}. The green shades correspond to the allowed regions extracted from the combination of most recent LHC (up to 13~TeV) and all Tevatron results. The yellow regions correspond to the combination of all most recent measurements with the expected future $W$-boson helicity fractions and $t$-channel single top quark production cross section measurements at the LHC, using 3000~fb$^{-1}$ of 14 TeV $pp$ data.

In Figure~\ref{fig:lim1} the limits obtained for the real part of the anomalous couplings are displayed, on the left plot, at 95\%~CL, where $\Vl = 1$ and all the other couplings are allowed to vary. Different measurements complement each other for narrowing down allowed regions. This is discussed in depth in~\cite{AguilarSaavedra:2011ct} where combining measurements leads to a better result, when compared to regions obtained from the individual measurements alone. Similarly, the right plot shows the allowed regions on the imaginary part of the anomalous couplings, where, once again, $\Vl = 1$ is fixed and the remaining couplings are allowed to vary. The shapes are slightly shifted from real counterparts, and are oriented differently in phase-space. But the key result is the same; a clear shrinking of the allowed region with the inclusion of each additional set of measurements. The three-dimensional nature of these plots enables a new representation which does not assume any constrain whatsoever on both real and imaginary parts of the anomalous couplings, and provides a new way to visualize the allowed values that anomalous couplings can hold, depending on each other. 
In Figure~\ref{fig:lim3}, two-dimensional plots (at 95\%~CL) are presented, with $\Vl$ fixed to unity and with all other couplings allowed to vary, as before. 
These two plots, when compared with the previous ones, make it easier to see exactly how these three-dimensional regions are made smaller with each measurement used. It is particularly visible how the inclusion of a possible future measurement of the single top quark production $t$-channel and $W$-boson helicity fractions at 14~TeV with high luminosity (yellow regions) would improve the limits by a significant factor, when compared with the results extracted from the combination of the current most precise measurements (green regions).

Figure~\ref{fig:lim4} shows three-dimensional limits obtained for real and imaginary parts of the two tensorial anomalous couplings, at 95\%~CL, assuming $\Vl = 1$, while all other couplings are allowed to vary. The left plot shows allowed regions for the real and imaginary parts of the right tensorial coupling, $\gr$, and the real component of the left tensorial coupling, $\gl$. The right plot shows allowed regions for the real and imaginary parts of $\gl$, and the real component of $\gr$.\footnote{The color scheme is the same as used in Figures~\ref{fig:lim1} and~\ref{fig:lim3}.} Similarly, Figure~\ref{fig:lim5} shows two-dimensional limits for the real and imaginary components of $\gr$, on the left, and $\gl$ on the right, at 95\%~CL, for $\Vl = 1$. As before, these plots make very clear how a future measurement of the single top quark production $t$-channel cross-section and $W$-boson helicity fractions, with high luminosity, at 14~TeV, would help constrain limits on anomalous contributions to the $Wtb$ vertex.

In addition to the two- and three-dimensional allowed regions, it is also useful to set limits on each anomalous contribution, while allowing all the other anomalous coupling to vary at the same time. These limits are presented in Table~\ref{tab:ReCoup}, at 95\%~confidence level, for the different considered scenarios. When compared with the previous most precise limits (up to 8~TeV), the current results (up to 13~TeV) show a significant improvement, larger than 10\%, in particular for the real component of $\gr$.

\begin{table}[h]
\begin{center}
\begin{tabular}{|c|c|c|c|}
\hline
               8~TeV            &        $\gr$    &       $\gl$        &     $\Vr$ \\ 
\hline
     Allowed Region ($Re$)   & [-0.15 , 0.09]  &    [-0.19 , 0.20]  & [-0.33 , 0.41] \\
     Allowed Region ($Im$)   & [-0.30 , 0.31]  &    [-0.20 , 0.19]  & [-0.37 , 0.38] \\
\hline
\hline
           13~TeV       &        $\gr$    &       $\gl$        &     $\Vr$ \\
\hline
     Allowed Region ($Re$)   & [-0.12 , 0.09]  &    [-0.17 , 0.18]  & [-0.31 , 0.38] \\
     Allowed Region ($Im$)    & [-0.26 , 0.27]  &    [-0.17 , 0.18]  & [-0.34 , 0.36] \\
\hline
\hline
           14~TeV       &        $\gr$    &       $\gl$        &     $\Vr$ \\
\hline
     Allowed Region ($Re$)   & [-0.07 , 0.07]  &    [-0.16 , 0.17]  & [-0.25 , 0.34] \\
     Allowed Region ($Im$)   & [-0.22 , 0.23]  &    [-0.16 , 0.17]  & [-0.32 , 0.31] \\
\hline
\end{tabular}
\caption{95\%~CL limits on the real and imaginary components of the anomalous couplings. These limits were extracted from the combination of $W$-boson helicities and single top quark production  
cross section measurements at LHC and Tevatron, up to 8~TeV (top), up to 13~TeV (center), and 14~TeV with high luminosity (bottom).}
\label{tab:ReCoup}
\end{center}
\end{table}
\section{Conclusions}

New limits were established on the anomalous contributions to the $Wtb$ vertex. These limits were calculated using the most precise measurements of the $W$-boson helicity 
fractions at the LHC, combined with measurements of single top quark production cross sections for different center-of-mass energies at the LHC and Tevatron. The allowed regions for both the real and imaginary components of the anomalous couplings were presented in three-dimensional plots, for different combinations of measurements, at 95\% CL, where all components were allowed to vary as free parameters in the global fit. The results presented in this paper showed an improvement larger than {10\%}, when compared with the previous most precise limits of new anomalous physics contributions to the $Wtb$ vertex. Prospected limits were also presented using the expected results of future measurements of the $t$-channel single top quark production cross-section and $W$-boson helicity fractions at 14~TeV, for a luminosity of 3000~fb$^{-1}$. The significant improvement observed motivates an important physics case for future single top quark production measurements at the LHC.


\section{Acknowledgements}


The authors would like to thank the Center for Theoretical Physics of the Physics Department at the New York City College of Technology, for providing computing power from their High-Performance Computing Cluster. The work of M.C.N.~Fiolhais was supported by FCT grant SFRH/BPD/100379/2014. The work of C.M.~Pease was partly supported by Macaulay Honors College. The authors would also like to thank Juan Antonio Aguilar-Saavedra and Nuno F. Castro for a long time collaboration.





\end{document}